\documentclass[10pt,twocolumn,letterpaper]{article}

\usepackage{wacv}
\usepackage{normalpkgs}
\usepackage[accsupp]{axessibility}  

\def\wacvPaperID{893} 

\wacvfinalcopy 

\ifwacvfinal
\fi

\ifwacvfinal
\usepackage[pagebackref=true,breaklinks=true,colorlinks,bookmarks=false]{hyperref}
\else
\usepackage[pagebackref=true,breaklinks=true,colorlinks,bookmarks=false]{hyperref}
\fi

\ifwacvfinal
\fi

\begin{document}

\title{Semi-Supervised Semantic Segmentation of Vessel Images using Leaking Perturbations}

\author{Jinyong Hou, Xuejie Ding, Jeremiah D. Deng  \\
Department of Information Science, University of Otago\\
{\tt\small robert.hou@postgrad.otago.ac.nz, \{emily.ding, jeremiah.deng\}@otago.ac.nz}}


\maketitle
\thispagestyle{empty}

\begin{abstract}
  Semantic segmentation based on deep learning methods can attain appealing accuracy provided large amounts of annotated samples. However, it remains a challenging task when only limited labelled data are available, which is especially common in medical imaging. In this paper, we propose to use Leaking GAN, a GAN-based semi-supervised architecture for retina vessel semantic segmentation. Our key idea is to pollute the discriminator by leaking information from the generator. This leads to more moderate generations that benefit the training of GAN. As a result, the unlabelled examples can be better utilized to boost the learning of the discriminator, which eventually leads to stronger classification performance. In addition, to overcome the variations in medical images, the mean-teacher mechanism is utilized as an auxiliary regularization of the discriminator. Further, we modify the focal loss to fit it as the consistency objective for mean-teacher regularizer. Extensive experiments demonstrate that the Leaking GAN framework achieves competitive performance compared to the state-of-the-art methods when evaluated on benchmark datasets including DRIVE, STARE and CHASE\_DB1, using as few as 8 labelled images in the semi-supervised setting. It also outperforms existing algorithms on cross-domain segmentation tasks.   
\end{abstract}


\section{Introduction} \label{sec:introduction}

Semantic segmentation on medical imaging has been an important and challenging visual recognition task aims at assigning a label from a set of classes to each pixel of the image~\cite{Chen2018a,Chen2019b,Majurski2019}. The segmentation techniques based on deep neural networks have been applied in various medical imaging modalities and attained the promising improvements. For fully supervised deep learning models, they rely on large volume of annotated datesets. However, for medical images, it is generally time-consuming and laborious for domain experts to label them especially for pixel-level.  

To overcome this limitation, semi-supervised learning methods are explored for medical image segmentation~\cite{Zhou2019a,Zhou2019,Bortsova2019}. But in real world, medical images are often taken from different scanners and environments, such as lights, angle, which alter the appearance of the images. Also, the images are related to the different categories of subjects (e.g. healthy or pathological), which varies the morphology of images. Both of them cause the domain shift for the different medical images. Towards these practical issues, we propose a model with generalization capacity which can transfer the knowledge from one domain to other unseen domains for the segmentation. 

Specifically, a GAN-based semi-supervised strategy~\cite{Salimans2016} is adopted as our basic framework. 
To improve its generalization, we add three improved strategies on the GAN-based framework. First, a U-Net style network is used as the discriminator due to its competitive performance in medical image segmentation~\cite{Ronneberger2015}. 
Secondly, we propose a polluted discriminator by introducing auxiliary ``leaking links'' from the generator to the discriminator, which force the generator to produce moderate (unrealistic) generations to  boost the semi-supervised learning performance. The polluted discriminator gains an improved learning ability due to the perturbation of probabilities during gradient optimization. Finally, the discriminator is regularized by the mean-teacher mechanism~\cite{Tarvainen2017}, which can improve the segmentation generalization because it is trained under input and weights perturbations. We also introduce a novel focal consistency penalty for the predictions of the student and teacher modules.  It can extend the training range when the two prediction results have significant shift.

Among various medical image modalities, retinal vessel images include rich information about pathological changes that are critical for early diagnosis and treatment of eye-related diseases~\cite{Yan2018}. Also, blood vessel analysis plays a fundamental role in different clinical fields~\cite{Moccia2018,Shin2019,Budai2013}. However, 
compared with other objects, the segmentation of vessels requires more fine-grained results, which is rather challenging~\cite{Fraz2012a}.  There are variations in terms of their sizes, shapes, and intensities in different local areas, forcing the segmentation model to learn the vessel's complex structure. Also, a micro-vessel can be as thin as just one to several pixels. 
On the other hand, there are a number of public-domain retinal vessels datasets, such as DRIVE~\cite{Staal2004}, STARE~\cite{Hoover2000}, and CHASE\_DB1~\cite{Fraz2012}. This provides a practical platform for us to utilize different datasets to evaluate our model's transfer learning~\cite{Pan2010} performance.


\section{Related Work} \label{sec:related_work}


\noindent
\textbf {Vessel Segmentation based on Deep Neural Networks:} 
Blood vessel segmentation is a hot topic in medical image analysis since the analysis results play an essential role in the diagnosis and intervention of many diseases~\cite{Shin2019}.  For retinal vessel images, it was pointed out in~\cite{Soomro2019} that the challenge in vessel segmentation is due to the unique complex structure of blood vessels. 
Deep convolutional neural networks applied to vessel segmentation have obtained some competitive results, and they allow a direct end-to-end function that can be trained using back propagation~\cite{Fu2016,Yan2018,Li2020d,Xu2020a,Kamran2021,Park2020}. Also, some models~\cite{Li2015,Fraz2012,Guo2019} can be used for cross-training evaluations, i.e., training on one dataset and testing on another dataset. It helps the framework be deployed to test fundus images obtained with different cameras and different environments. These successes, however, rely on ample annotated data, which usually is not the case for medical images as labelling full images is time-consuming and expensive. The limited availability of annotated vessels remains the bottleneck for deep neural networks' application in real-world vessel segmentation tasks. 

\noindent
\textbf {Semi-Supervised Segmentation:} 
To address the data limitation, several semi-supervised semantic segmentation methods have been proposed for medical images~\cite{Zhou2019,Bortsova2019,Zhou2019a}. The unlabelled data can be used freely to improve fully supervised performance. The generative adversarial network (GAN) has shown potential performance on semi-supervised learning~\cite{Hong2015, Salimans2016, Gan2017}. The discriminator also serves as a classifier, not only discriminates real examples from fake ones but also classifies the former~\cite {Salimans2016}. 
Using the GAN, the generator can create extensive realistic visual data, enabling the discriminator to learn better features for more accurate pixel classification in semantic segmentation~\cite{Souly2017}. It has been pointed out in~~\cite{Dai2017, Kumar2017} that good semi-supervised classification performance and perfect generators could not be obtained simultaneously in the GAN-based methods. Further insights were provided in~\cite{Kumar2017} on how fake examples influence semi-supervised learning, and that only moderate fake examples can its boost performance. Inspired by these works, we propose a novel GAN method that can generate moderate samples and force the discriminator to better learn the fine-grained vessel structure. 

Another semi-supervised state-of-the-art technique is Mean-Teacher, a method that averages model weights to form a target-generating teacher model~\cite{Tarvainen2017,Athiwaratkun2019}. The mean-teacher paradigm is applied in cross-domain detection~\cite{Cai2019}, where the domain gap was bridged with consistency regularization in a teacher-student scheme. In our work, the mean-teacher is explored to improve the segmentation generalization capacity by the perturbation from inputs and weights in a teacher-student scheme. Meanwhile, we also propose a focal consistency to penalize the teacher and student networks' predictions. The novel consistency makes the network focus on the uncertain pixels, especially for the thin vessel, challenging to distinguish. 

\noindent
\textbf{Skip Connection in Deep Learning:} The use of skip connections often takes a crucial role in deep neural networks classifiers~\cite{He2016,Lee2015,Huang2017,Srivastava2015b} as it can effectively mitigate the vanishing gradient problem.
Another research line combines the shallow information from lower layers with deep information from high layers
~\cite{Ronneberger2015, LongJ2015, Rasmus2015a} and hence leads to more accurate recognition results. 
In~\cite{Shama2019}, the authors propagate the discriminator information back to the generator, utilizing an iterative feedback loop to improve the generator's performance. Reversely, in our work, we employ skip connections to `leak' information from the generator to the discriminator to boost the performance of semi-supervised learning.

\section{Proposed Model} \label{sec:model}


\begin{figure*}[thbp] 
    \centering
    \includegraphics[width=0.7\textwidth]{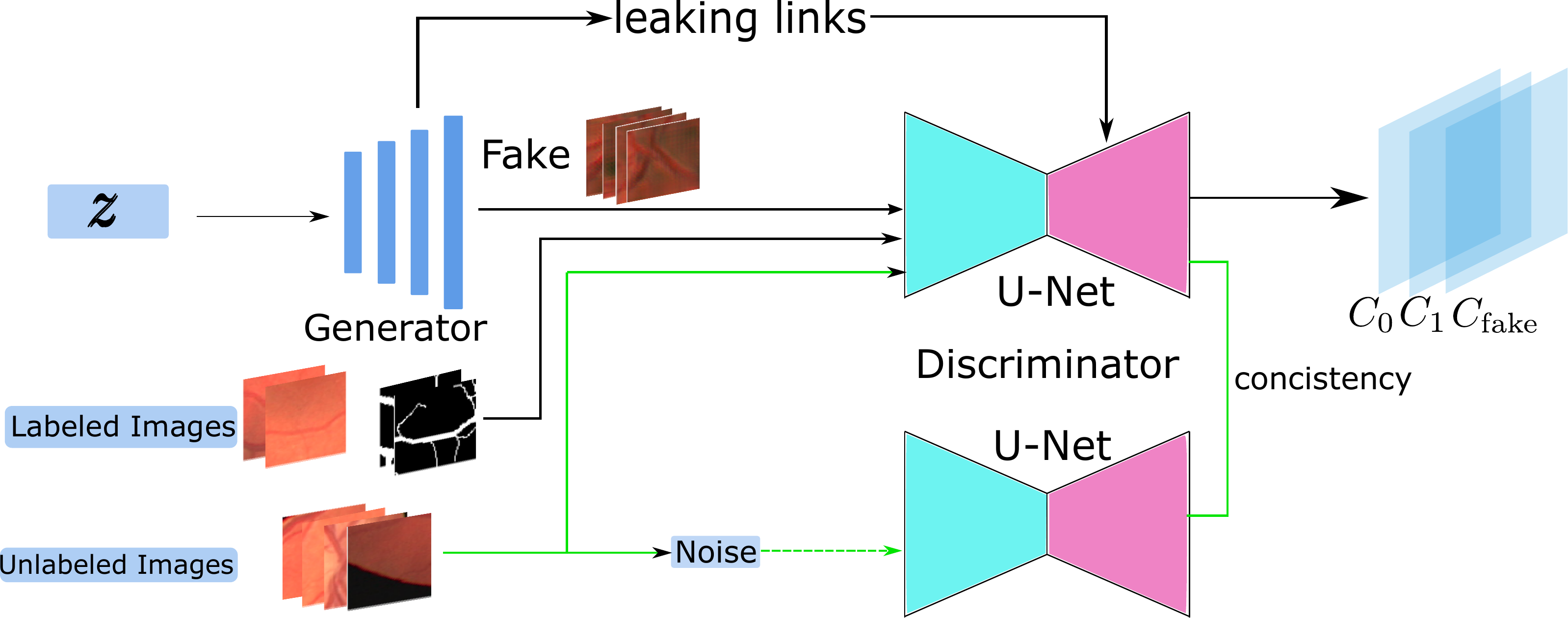}
    \caption{Model Schematic for retina semantic segmentation. There are three inputs (labelled, unlabelled and fake data) to feed into the discriminator. U-Net is utilized as the discriminator in the proposed model. During training, unlabelled data are also used to train a teacher discriminator.}
    \label{fig:model_sche}
\end{figure*}


\subsection {Problem setting}
\label {subsec:problem setting}

We have a training set $\mathcal{D}_s$ that contains labelled samples $\{(\bm x_{l}^i, y^i)|i=1,\cdots,n\}$ and unlabelled samples $\{\bm x_\mathrm{ul}\}$,
where each sample $\bm x_{l}, \bm x_{\text{ul}}\in \mathcal{\bm X}$, only limited samples $\bm x_{l}^{i}$ are associated with a label $y_i\in \mathcal{\bm Y}$. $\mathcal{\bm X}$ is the feature space, such as an $m$ dimensional space of real numbers $\mathbb{R}^m$, and $\mathcal{\bm Y}$ is the label space. The unlabelled data $\bm x_{\text{ul}}$ is used to improve the predictions of the classifier. The classifier can provide labels for unlabelled data. 

Our goal is to design a generalization segmentation framework to achieve good performance on semi-supervised learning scenarios. It can also perform well in cross-training scenarios, i.e., training on dataset $\mathcal{D}_s$ and testing on dataset $\mathcal{D}_{t}$. Here the $\bm x_t, \bm x_{l}, \bm x_{\text{ul}}\in \mathcal{X}$ are in the same feature space. However, there is domain shift between $\mathcal{D}_s$ and $\mathcal{D}_t$, and therefore different distributions $p(\mathbf{X}_s) \neq p(\mathbf{X}_t)$. In the real world, the distribution discrepancy between training data and testing data will often hinder performance. Therefore, generalization segmentation networks are designed to be transferred to other domains (even unseen scenarios), and it performs well for domain shift.

\subsection {Model framework}
\label{subsec:model_framework}

The proposed framework is shown in Fig.~\ref{fig:model_sche}. Our model follows the GAN-based semi-supervised approach. We aim to classify each pixel of the retinal image to one of the $K$ classes. 

In the model, the fake image $\bm x_f$ is labelled with an additional $K+1$ class, and $p_{k}(y=K+1|\bm x_f)$ can be used to supply the probability that data $\bm x_f$ is fake. During the training, the unlabelled data $\bm x_{\text{ul}}$ has to be one of the $K$ classes. The generated fake samples can be used with unlabelled data to explore the hidden data structure to support semantic segmentation. The labelled data $\bm x_l$ is utilized for training the discriminator to recognize the pixel's categories. 
Therefore, the discriminator's loss function is given as a sum of supervised and unsupervised parts:

\begin{equation}  \label{eq:semi-supervised}
  \mathcal {L} = \mathcal {L}_{\text{sup}} +\mathcal {L}_{\text{unsup}}.
\end{equation}

The loss for supervised images is a cross-entropy loss between labels and probabilities predicted by the discriminator, as used in standard segmentation networks:
\begin{equation}  \label{eq:GAN_based_supervised_objective}
  \mathcal {L}_{\text{sup}} = -\mathbb{E}_{\bm x_l,y\sim p_{\text{data}}(\bm x_l,y)}\log p_k(y|\bm x_l,y<K+1).  
\end{equation}

The loss for unsupervised images is defined as
\begin{equation}  \label{eq:GAN_based_unsupervised_objective_GAN}
  \resizebox{0.3\textwidth}{!}{$
  \begin{aligned}
    \mathcal{L}_{\text{unsup}} &= -\mathbb{E}_{\bm x_{\text{ul}} \sim p_{\text{data}}(\bm x_{\text{ul}})}\log D(\bm x_{\text{ul}})\\
  &-\mathbb{E}_{\bm z \sim G} \log(1-D(G(\bm z)),
\end{aligned}$}
\end{equation}
where $\bm z$ is the noise used to generate the fake images. Due to the over-parameterization of the $K+1$ outputs in the discriminator, the $(K+1)$-th logit, $l_{K+1}(\bm x_f)$ can be clamped to 0 for all the $\bm x_f$~\cite{Salimans2016}. Then the discriminator $D$ can be obtained by $D(\cdot) = \frac{Z(\cdot)}{Z(\cdot) + 1}$
for unlabelled and fake images, where $Z(\cdot) = \sum_{i=1}^{K}\exp(l_{i}(\cdot))$. Also, the feature matching loss has a better performance to train the model.

In the model, the \textit{discriminator} also functions as a classifier for semantic segmentation. It is crucial for the segmentation performance. Among various discriminator models that can be chosen for the task, we choose \textit{U-Net} as it is a good performer on medical image segmentation.

\begin{figure}[thbp]
  \centering
  \includegraphics[width=0.5\textwidth]{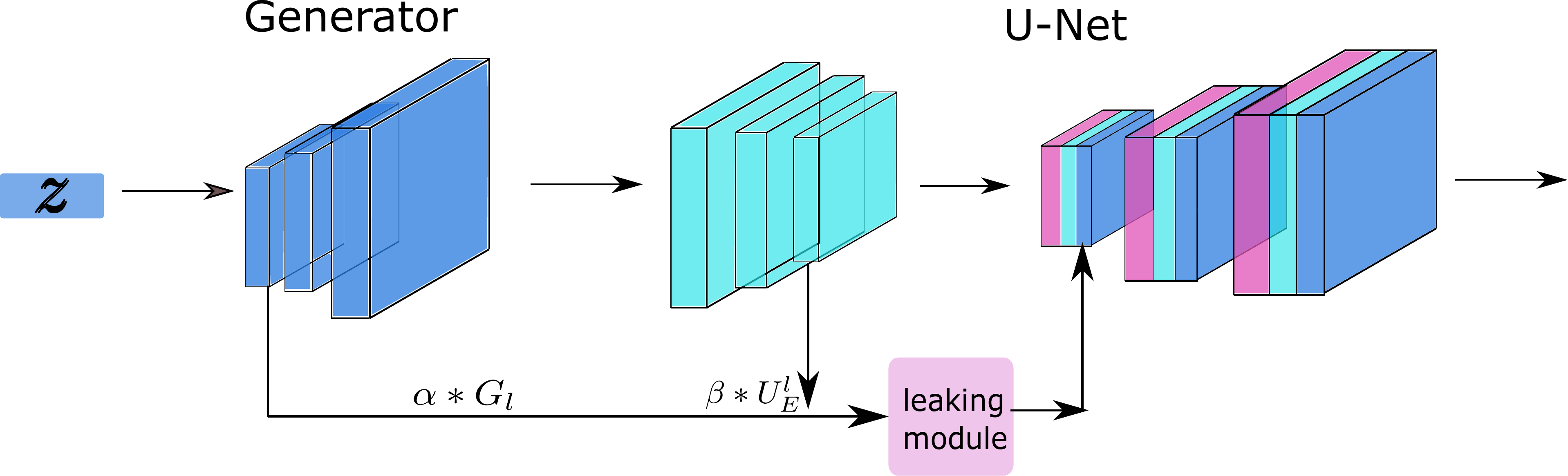}
  \caption{\small For the information leaking module, the outputs of the generator layers are concatenated with the upsampling parts of U-Net.}
  \label{fig:leaking_gan_unit}
\end{figure}

To boost the semi-supervised learning, we propose a new leaking mechanism for the discriminator, as shown in Fig.~\ref{fig:leaking_gan_unit}. We add a skip connection between the generator and the discriminator. Let us denote the $l$-th layer outputs of the generator, the contracting path and expanding path of the U-Net by  $\{G^l\}_{l=1}^{n-1}$,$\{U_c^l\}_{l=n-1}^1$ and $\{U_e^l\}_{l=n+1}^{2n-1}$ respectively, where $n$ is the index number of the bottom layer in U-Net. The activation of the generator and the U-Net contracting path are concatenated by the \textit{information leaking} module. We denote the $l$-th output layer of information leaking module as $\{L ^l\}_{l=1}^n$. The activation of each layer in the information leaking module can be described as 
\begin{equation}
  L ^l= [\alpha^l G^l, \beta ^l U_c^l],
\end{equation} 
where $[\cdot, \cdot]$ denotes the concatenated operation. The coefficients $\alpha ^l$ and  $\beta ^l$ are scaling parameters for the outputs in the layer.  Then the output of the information leaking module is added to the corresponding activation of the expanding path, as follows: 
\begin{equation}
  U_e^l= f [U_e^l, \gamma L ^l],
\end{equation} 
where $f$ is the activation function, and $\gamma$ is a switch function.  When $\gamma=(1, 1)$ the information leaking module is activated; when $\gamma=(0, 1)$, the module is disabled. Thus, the information leaking module can be flexibly switched on and off. 

Then the discriminator gives the decision according to the features polluted by leaking information from the generator. It leads to generate unrealistic examples with a perfect generator. The unrealistic examples can boost the semi-supervised learning performance, as proven in~\cite{Kumar2017, Dai2017}. However, different from bad GAN~\cite{Dai2017}, our generator is perfect. The unrealistic examples are generated due to the polluted discriminator. Also, the polluted features induce the perturbation on the probability during the gradient optimization. These kinds of perturbations also improve the learning ability of discriminator (U-Nets). A detailed explanation of the leaking module is in section~\ref{subsec:leaking_gan_explanation}.  

With the information leaking strategy, the proposed model tends to leverage the unlabelled data to explore the more intrinsic properties of data and achieve better results. 
However, variations factors such as image intensity changes, noise, and low contrast, still present significant obstacles for the segmentation model~\cite{Moccia2018}. 

To further improve the discriminator's generalization over the variations in medical images, we adopt the mean-teacher paradigm~\cite{Tarvainen2017,Cai2019} as an additional regularization. 
Specifically, follow the protocol of~\cite{Cai2019}, we only apply the noise to the teacher path. Also, under the medical image scenarios, we apply the added noise perturbations but not the augmentation for this. The student's inputs $\bm x_{\text{ul}}^s$ polluted by noise are fed into the teacher network. That is, for the teacher's network, inputs $\bm x_{\text{ul}}^t$ become  $\bm x_{\text{ul}}^t=\bm x_{\text{ul}}^s+\lambda \epsilon$, with $\epsilon \sim \mathcal{N}(0,I)$, $\lambda < 1$. Another perturbation is from the weights. Here we notify the weights of student discriminator $w_{D_s}$. With the same architecture as the students, the teacher discriminator is parameterized by $w_{D_t}$. According to the mean-teacher strategy~\cite{Tarvainen2017}, the teacher network weights $w_{D_t}$ are the exponential moving average (EMA) of the student weights:
\begin{equation}
\omega_{D_t}^\tau=\alpha  \omega_{D_t}^{\tau-1}+(1-\alpha)\omega_{D_s}^{\tau},
\end{equation}
where the $\alpha$ is the smoothing coefficient parameter that controls the updating of teacher weights, which is usually set between 0.9 and 0.999, and the $\tau$ is the training step. Then a consistency loss~\cite{Tarvainen2017} is employed to penalize the difference between the student's prediction $f_{s}(\bm x_{ul}^{s};\omega_{D_{s}})$ and the teacher's $f_{t}(\bm x_{ul}^{t};\omega_{D_{t}})$:
\begin{equation} 
  \mathcal{L}_{\text{cons}}(\bm x_{\text{ul}})=||f_s(\bm x_{\text{ul}}^{s};\omega_{D_s})-f_t(\bm x_{\text{ul}}^{t};\omega_{D_t})||.
\end{equation}


\subsection {Probability perturbation analysis}
\label{subsec:leaking_gan_explanation}

For the GAN-based semi-supervised approach, it has been pointed out that a moderate generation can boost the performance~\cite{Kumar2017}\cite{Dai2017}.
To obtain a moderate generation, a complementary generator (i.e., a bad generator) was introduced in~\cite{Dai2017} to improve the generalization of a semi-supervised learning framework. In our method, we keep the perfect generator to capture the data distribution but introduce a ``polluted" discriminator for moderate generations. 


\begin{definition}[Polluted Discriminator]
  The polluted discriminator distinguishes the noise-added example $\tilde {\bm x}$ 
  from the generated example $\bm x_g$. The noise is leaked from the generator by the information leaking module. Then for any optimal solution $V(G,D)$,  $p_G \approx p_d$.
\begin{equation}
  \resizebox{0.49\textwidth}{!}{$
  \min \limits_{G} \max \limits_{D} V(D,G) = \mathbb {E}_{\bm x\sim p_{\textnormal{data}}}[\log D(\tilde{\bm x})] + \mathbb{E}_{\bm z\sim p(\bm z)}[\log (1-D(G(\bm z)))].$}
\end{equation}
\end{definition}

The polluted discriminator leads the generator to a distorted version of the real example; it pushes the generator away from producing strong generations. Furthermore, the distance between the distorted image $\tilde{\bm x}_{\text{ul}}$ and the real one $\bm x_{\text{ul}}$ can be controlled by the leaking information module. That means we can control how much the distortion the generations have. 


Following the approach in~\cite{Kumar2017}, we now give an analysis on the probability perturbation effects of information leaking and illustrate why it works. The unsupervised objective in Eq.~(\ref{eq:GAN_based_unsupervised_objective_GAN}) can be rewritten as follows:
\begin{equation}
  \resizebox{0.35\textwidth}{!}{$
  \begin{aligned}
    \mathcal{L}_{\text{unsup}} &= - \mathbb{E}_{\bm x_f \sim p_{\text{fake}}} \log \frac{\exp(l_{K+1}(\widetilde {\bm x}_f))}{\sum_{i=1}^{K+1} \exp(l_{i}(\widetilde {\bm x}_f))}\\   
    &-\mathbb{E}_{\bm x_{\text{ul}} \sim p_{\text{data}}}\left[ \log (1-\frac{\exp(l_{K+1}(\widetilde{\bm x}_{\text{ul}}))}{\sum_{i=1}^{K+1}\exp(l_{i}(\widetilde {\bm x}_{\text{ul}}))}) \right],
  \end{aligned} $}
\end{equation}
where $p_{\text{model}}(y=K+1|\bm x)$ is replaced by $\frac{\exp (l_{K+1}(\bm x))}{\sum _{i=1}^{K+1}\exp(l_{i}(\bm x))}$.
When we take the derivative w.r.t the weights of the discriminator, we get
\begin{equation} \label{eq:noised_gradient}
  \resizebox{0.48\textwidth}{!}{$
  \begin{aligned}
    \nabla \mathcal{L}_{\text{unsup}} 
    &= \mathbb{E}_{\bm x_f \sim p_{\text{fake}}}\sum _{i=1}^{K} \underbrace{p_{\text{model}}(y=i|\bm x_f)\nabla l_{i}(\bm x _f)}_{a_i(\tilde{\bm x}_f)} \\
    & -\mathbb{E}_{\widetilde{\bm x}_{\text{ul}}\sim p_{\text{data}}} \sum_{i=1}^{K}[\underbrace{p_{\text{model}}(y=i|\widetilde{\bm x}_{\text{ul}}) p_{\text{model}}(y=K+1|\widetilde{\bm x}_{\text{ul}})] \nabla l_i(\widetilde{\bm x}_{\text{ul}})}_{b_i(\widetilde{\bm x}_{\text{ul}})}].
  \end{aligned}$}
\end{equation}

Our model leaks the information from the generator to the discriminator, concatenating as a feature vector to the expanding path. This makes the discriminator give the decision on $\tilde{\bm x}_{\text{ul}}$ instead of the real input $\bm x_{\text{ul}}$. Under this situation, we assume there is some perturbation on the probability in the $ b_{i}(\widetilde{\bm x}_{\text{ul}})$ part, with the probability of being 
classified as normal classes reduced, and the probability of being classified as ``fake'' increased, all due to the pollution resulted from the leaking link. It becomes
\begin{equation} \label{eq:delta_unsup_leaking}
  \resizebox{0.45\textwidth}{!}{$
  \begin{aligned}
    b_{i}(\widetilde {\bm x}_{\text{ul}}) = -\mathbb{E}_{{\bm x}_{\text{ul}}\sim p_{\text{data}}}\sum_{i=1}^{K}[&(p_{\text{model}}(y=i|{\bm x}_{\text{ul}})- \varepsilon_1)\\
    &(p_{\text{model}}(y=K+1|{\bm x}_{\text{ul}})+\varepsilon_2)] \nabla l_i({\bm x}_{\text{ul}}).
    \end{aligned}$}
\end{equation}

The implication of Eq.~(\ref{eq:delta_unsup_leaking}) can be analysed as follows.
\begin{enumerate}
  \item The perturbation can increase the gradient at the early training. At the beginning of training, the examples generated from the generator are very weak and easy to distinguish from the real examples. There exists $i\le K$, $p(y=i|{\bm x}_{\text{ul}})\approx 1$, $p(y=K+1|\bm x_{\text{ul}})\approx 0$, and $p(y=i|x_f)\approx 0$. Under this situation, there is no gradient flow from Eq.~(\ref{eq:delta_unsup_leaking}), i.e., zero contribution for unsupervised part in the overall gradient. This issue is remedied when the leaking information is added to pollute the unlabelled feature. As a result, the $p(y=K+1|{\bm x}_{\text{ul}})\approx 0+\varepsilon_2$ and $p(y=i|\bm x_f)\approx 1- \varepsilon_1$, where $\varepsilon_1, \varepsilon_2 >0$, so that the unlabelled data contribute to the gradient.

  \item The perturbation also tends to prolong the training of the discriminator. For a perfect GAN, with the training going on, the generator captures the data distribution. The discriminator can not distinguish the two distributions, which means $p_{\bm x_f}=p_{\bm x_{\text{ul}}}$, and the discriminator probability becomes $D(\bm x)\approx\frac{1}{2}$~\cite{Goodfellow2014}. When the discriminator is polluted by information leaking, $p(y=i|\bm x_{\text{ul}})$ potentially decreases, and $p(y=K+1|\bm x_f)$ increases: $p(y=i|\bm x_{\text{ul}},y\le K)\approx \frac{1}{2}- \varepsilon_1$ and $p(y=K+1|\bm x_f)\approx \frac{1}{2}+\varepsilon_2$, where $\varepsilon_1, \varepsilon_2>0$ decrease gradually with the training going on. Thus the discriminator gets more opportunities to learn better feature representations.
\end{enumerate}


\subsection {Focal consistency} 
\label{subsec:focal_consistency}

In retina vessel images, it is challenging to deal with micro-vessels with delicate structures as they may take up only a few pixels in width and the classification can be easily disrupted by noise. There is a severe class imbalance problem during training. Inspired by~\cite{Lin2017a}, we propose a focal consistency loss function  
which down-scales the weights of easy examples, and thus guides the training to focus on the hard negatives. 

The focal loss function~\cite{Lin2017a} tries to mitigate the effect of the cross entropy loss being overwhelmed by large class imbalance. Let $p_t$ denote the probability for class $t$. The relevant cross entropy $-\log p_t$ can be scaled by a weighting factor $\alpha_t$ (which can be set as the inverse of class frequency or as a hyperparameter). Adding a modulating factor $(1-p_t)^\rho$, with $\rho>0$ being a tunable focusing parameter, we have the balanced focal loss function defined as
\begin{equation}
  \text{FL}(p_t)=-\alpha_t(1-p_t)^\rho \log p_t.
\end{equation}

Based on this, we propose a novel \textit{focal consistency loss function} as follows by
considering the prediction difference between the student and teacher networks:
\begin{equation}  \label{eq:focal_consistency}
  \mathcal{L}_{\text{cons-FL}}=-\alpha_t(|p_t(\bm x_{\text{ul}})-p_t(\bm x_{\text{ul}}')|)^\rho [\log p_t(\bm x_{\text{ul}})-\log p_t(\bm x_{\text{ul}}')]
\end{equation}
where $|p_t(\bm x_{\text{ul}})-p_t(\bm x_{\text{ul}}')|$ is the perturbation factor. 
$\log p_t(\bm x_{\text{ul}})-\log p_t(\bm x_{\text{ul}}') $ is the prediction distance between 
the teacher and the student networks where the input signal and weights happened to perturbation. The distance of the two predictions decides the perturbation factor. It modulates the training range according to the consistency loss. When an example is aligned not well, the weight is improved due to the larger difference. When the two predictions are similar, it downs the loss weight. As the same as the reference~\cite{Lin2017a}, the parameter adjusts the rate at which aligned-well examples are down-weighted.

The consistency loss can extend the training range when the two prediction results have a more massive shift. It overcomes the distribution mismatch, possibly caused by the different view angles, lighting conditions, or acquisition devices. The dynamically scaled predict discrepancy boosts the generalization of the discriminator to adapt the transfer learning tasks.

\subsection{Learning in LeakGAN}
\label{subsec:learning_in_LeakGAN}

Having illustrated our framework with all loss functions defined, we give the overall learning objective as follows:
\begin{equation}  \label{eq:overall_objective_LeakGAN}
  \mathcal{L}^* = \min_G \max_D (\lambda_1 \mathcal{L}_{\text{sup}} + \lambda_2 \mathcal{L}_{\text{unsup}} + \lambda_3 \mathcal{L}_{\text{cons-FL}})
\end{equation}
where $\lambda_1, \lambda_2, \lambda_3$ are the trade-off hyper-parameters to control the  regularization of the model weights. Our model is trained like a standard GAN. The discriminator (also a classifier) is optimized by the joint objective. According to the analysis in section~\ref{subsec:focal_consistency}, we also apply the focal loss to $\mathcal{L}_{\text{sup}}$ to tackle the imbalance problem during the training. Meanwhile, for the optimization of the  generator, we observe that our model can be trained by the standard GAN's protocol empirically:
\begin{equation} \label{eq:LeakGAN_adversarial_objective}
  \mathcal{L}_{\text{adv-}} =  \mathbb{E}_{\bm z \sim p(\bm z)} \log (1 - p(D(G(\bm z)))).
\end{equation}

During the training, we optimize the models iteratively. Besides, the consistency of the mean-teacher regularizes the discriminator as its student network. 

\section{Experiments} \label{sec:experiments}


\begin{table*}[!htbp]
    \centering
    \caption{\small Performance evaluation for DRIVE, STARE and CHASE\_DB1. Semantic segmentation accuracy (Acc), Sensitivity (Se), and Specificity (Sp) are considered for the evaluations. Best performance highlighted in bold.}
    \resizebox{0.75\textwidth}{!}{
        \begin{tabular}{|c|c|c|c|c|c|c|c|c|c|}
            \hline
            Dataset & \multicolumn{3}{|c|}{DRIVE} & \multicolumn{3}{|c|}{STARE} & \multicolumn{3}{|c|}{CHASE\_DB1}   \\
            \hline
            Methods & Acc(\%)  & Sp(\%) & Se(\%) & Acc(\%)  & Sp(\%) & Se(\%) & Acc(\%) & Sp(\%) & Se(\%) \\
            \hline \hline
            MS-NFN~\cite{Wu2018}  & 95.67  & 98.19  & 78.44  & - & - & - & 96.37   & \textbf{98.47}  & 75.38  \\ 
            DeepVessel~\cite{Fu2016}  & 95.23  & - & 76.03  &  95.85  & - & 74.12  & 94.89 & - & 71.30 \\
            SegmentLoss~\cite{Yan2018} & 95.42  & 98.18  & 76.53  &  96.12  & 98.46 & 75.81 & 96.10 & 98.09 & 76.33 \\ 
            IterNet~\cite{Li2020d} & 95.73 & \textbf{98.38} & 77.35 & \textbf{97.01} & \textbf{98.86} & 77.15 & 96.55 & 98.23 & 79.70 \\
            \hline 
            LeakGAN 3  & 93.31    & 88.52  & 94.31  & 94.83 & 95.19 & 84.58 & 94.20  & 80.78  & 94.25  \\
            LeakGAN 5  & 94.69    & 83.14  & 97.00  & 94.89 &  88.21  & \textbf{96.32}  & 96.65   & 92.03  & 92.25  \\
            LeakGAN 8 & \textbf{95.74}  & 86.72  & \textbf{97.50}  &  95.65 &  91.86 & 91.02& \textbf{96.83}   & 92.21  & \textbf{94.72} \\ 
            \hline  
    \end{tabular}}
    \label{table:performance}
\end{table*}

\subsection {Datasets}
\label{subsec:dataset} 

The proposed method is evaluated on three public datasets, DRIVE~\cite{Staal2004}, STARE~\cite{Hoover2000} and {CHASE\_DB1}~\cite{Fraz2012}. These datasets provide corresponding vessel segmentation manually labelled by domain experts, which can be seen as the ground truth segmentation. The details of the datasets are as follows. 

\noindent
\textbf{DRIVE:} There are 40 color fundus images with a resolution of $565 \times 584$ pixels. These images were captured by a Canon CR5 non-mydriatic camera with a 45-degree field of view (FOV). The bit-depth is 8-bit. 

\noindent
\textbf{STARE:} It consists of 20 color fundus images, with half of them containing signs of various pathologies. These images were captured with the resolution of $700 \times 605$ pixels by a canon TopCon TRV-50 fundus camera with 35-degree FOV, and the bit-depth is 8-bit. 

\noindent
\textbf{CHASE\_DB1:} There are 28 color fundus images from patients who participated in the child heart and health study in England.  These images were captured with the resolution of $ 999 \times 960$ pixels by a Nidek NM 200D fundus camera with 30-degree FOV, and the bit-depth is 8-bit.

\subsection {Quantitative Results}
\label{subsec:quanti}

We evaluate the performance of our model on the three datasets and compare our model with state-of-the-art methods such as DeepVessel~\cite{Fu2016}, SegmentLoss~\cite{Yan2018}, MS-NFN~\cite{Wu2018} and IterNet~\cite{Li2020d}. Followed the protocol in~\cite{Soomro2019}, we use segmentation accuracy (Acc), specificity (Sp), sensitivity (Se) as quantitative results to measure the segmentation performance. 
They are defined as follows.
\begin{equation*}  \label{eq:ch5_se}
  \text{Se} = \frac{\text{TP}}{\text{TP + FN}},
\end{equation*}
\begin{equation*}  \label{eq:ch5_sp}
  \text{Sp} = \frac{\text{TN}}{\text{TN + FP}},
\end{equation*}
\begin{equation*}  \label{eq:ch5_acc}
  \text{Acc} = \frac{\text{TP + TN}}{\text{TP + FN + TN + FP}},
\end{equation*} 
where $\text{TP}$, true positives, are the vessel pixels predicted correctly. FP (false positive), TN (true negative), and FN (false negative) are similarily defined. 

During the evaluations, the trained patches are cropped randomly. To evaluate the semi-supervised learning performance using different amounts of labelled data, we set the number of labelled training retina image to be 3, 5, and 8, respectively. The rest of the training images are treated as unlabelled samples. The evaluation result is shown in Table~\ref{table:performance}.
  
As seen, when using 3, 5 labelled images only (``LeakGAN 3'' and ``LeakGAN 5''), the performance of LeakGAN is already close to the comparison methods. When 8 labelled images are used (``LeakGAN 8''), its performance surpasses that of the counterparts with full supervised strategy on most of the criteria. Furthermore, our model has a balanced performance for the $\text{Se}$ and $\text{Sp}$, while the results could benefit from a higher Se value. The results show performance boost for all the three different tasks. It demonstrates that the unlabelled images can be effectively employed to train good classification boundaries, whereas more labelled images are beneficial for improved performance. 

\subsection {Qualitative Results}
\label{subsec:visual}

\begin{figure*}[!thbp]
    \centering
        \centering
        \includegraphics[width=0.85\textwidth]{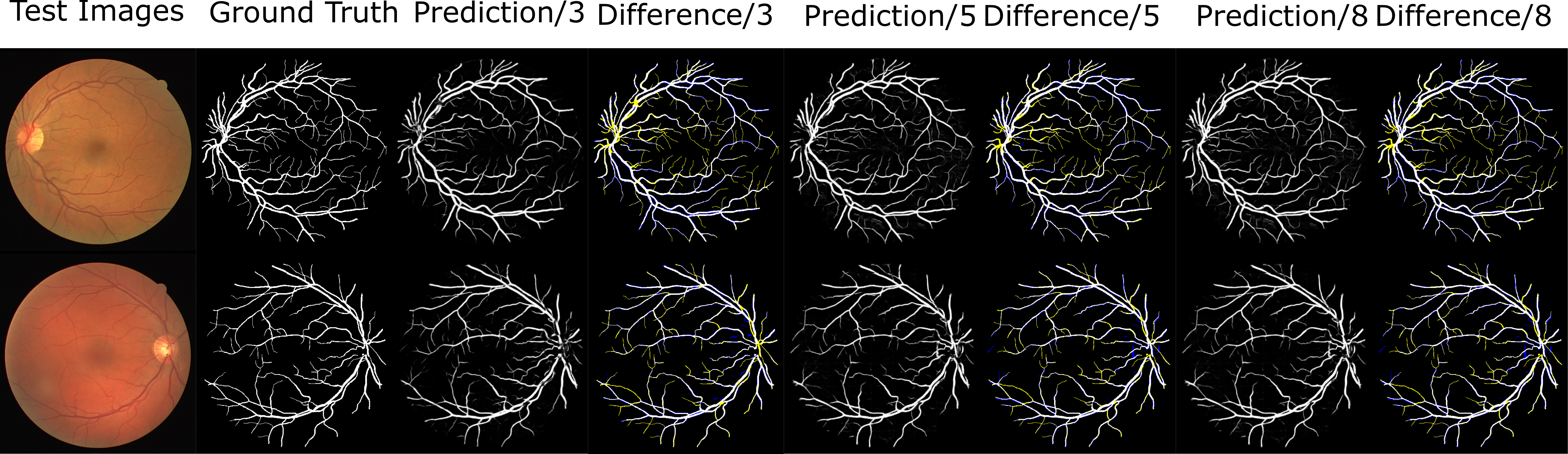}
    \caption{Visualization of the semantic segmentation on testing images for DRIVE. The results are shown when 3, 5 and 8 labelled training images are used. For the differences, the blue pixels being false positive, and yellow being false negative.}
    \label{fig:vis_entire}
\end{figure*}

We proceed with the visual assessment of the semantic segmentation results. First,  for DRIVE, whole-image predictions are shown in Fig.~\ref{fig:vis_entire}. Next, for more detailed visualization, patch predictions are also shown in Fig.~\ref{fig:vis_patches}. The results of the STARE and CHASE\_DB1 are shown in Supplementary Fig.~\ref{fig:vis_entire_supp} and Fig.~\ref{fig:vis_patches_supp}.

From the figures, we can see that our model predicts the retina fundus images well under the semi-supervised settings. They keep the preliminary information such as the retina's shape and the vessels' directions, though there are some differences. The differences in blue show the false positive pixels and yellow for false-negative ones. The blue pixels are mainly because of the extra vessel prediction alongside the real vessel, and the yellow pixels are due to a weak prediction value compared with the ground truth.

From the detailed patches prediction in Fig.~\ref{fig:vis_patches}, we can see that the thin vessels can be accurately detected from these three retina datasets. For example, in the third column in Fig.~\ref{fig:vis_patches}, the thin structure of the vessel is presented well by the prediction. 

\begin{figure}[!thbp]
        \centering
        \includegraphics[width=0.35\textwidth]{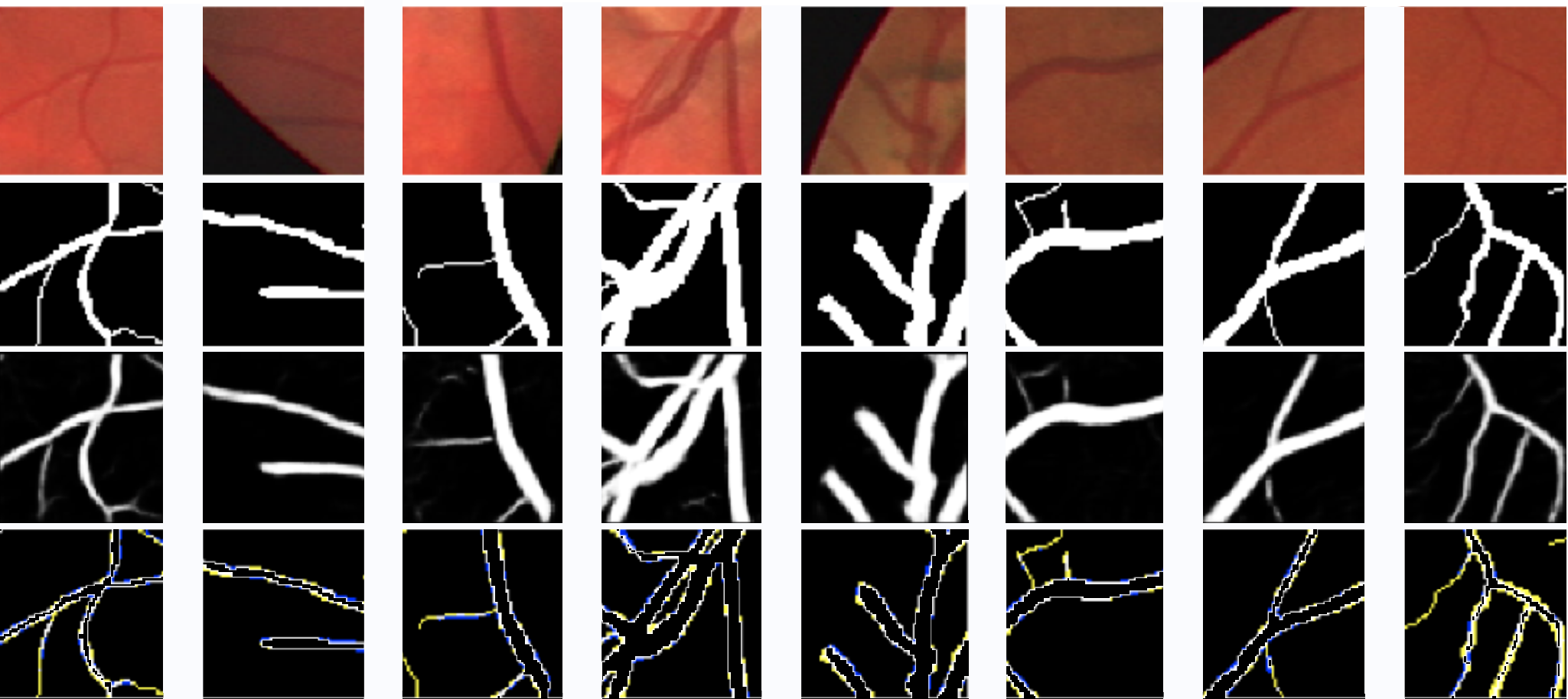}
    \caption{Patched segmentation results for the scenario of 8 labelled training image using DRIVE. The testing patches and their ground truth are in first and second row respectively. The third row is for the predictions. The differences are shown in the last row, blue pixels being false positive, and yellow being false negative.}
    \label{fig:vis_patches}
\end{figure}

\subsection {Results of Cross-Domain Semantic Segmentation}
\label{subsec:cross-domain}

To evaluate our framework's generalization capacity, we conduct cross-domain semantic segmentation experiments following Algorithm ~\ref{algo:LeakGAN_algorithm_cross_domain} (see Supplementary). The proposed model is trained on one dataset and tested on another dataset. 
All the evaluations are under 8-labelled images setting. The results are shown in Table~\ref{table:cross_performance}. In general, our model can generalize well to other fundus images captured with different cameras. The individual cross-dataset results are a little lower than the performance for within-dataset training and testing. Also, the performance from STARE or CHASE\_DB1 to DRIVE is better than the opposite transfer performance. It might be due to the higher resolutions of STARE and CHASE\_DB1. 

\begin{table*}[!thbp]
    \centering
    \caption{\small Performance of cross-domain evaluation. Semantic segmentation accuracy (Acc), Sensitivity (Se), and Specificity (Sp) are considered for the evaluations} 
    \resizebox{1.0\textwidth}{!}{
      \begin{tabular}{|c|c|c|c|c|c|c|c|c|c|c|c|c|c|c|c|c|c|c|}
            \hline
            Training & \multicolumn{6}{|c|}{DRIVE} & \multicolumn{6}{|c|}{STARE} & \multicolumn{6}{|c|}{CHASE\_DB1}   \\
            \hline
            Testing & \multicolumn{3}{|c|}{STARE} &\multicolumn{3}{|c|}{CHASE\_DB1} & \multicolumn{3}{|c|}{DRIVE} & \multicolumn{3}{|c|}{CHASE\_DB1} & \multicolumn{3}{|c|}{DRIVE} & \multicolumn{3}{|c|}{STARE} \\
            \hline
            Methods & Acc(\%)  & Sp(\%) & Se(\%) & Acc(\%)  & Sp(\%) & Se(\%) & Acc(\%) & Sp(\%) & Se(\%) & Acc(\%)  & Sp(\%) & Se(\%) & Acc(\%)  & Sp(\%) & Se(\%) & Acc(\%) & Sp(\%) & Se(\%) \\
            \hline \hline 
            EnsembleModel~\cite{Fraz2012}  & 94.95 & 97.70  & 70.10 & - & - & - & 94.56 & 97.92 & 72.42 & 94.15 & 96.65 & 71.03  & - & - & - & - &  -  & -  \\ 
            Cross-Mod~\cite{Li2015}  & 95.45  & \textbf{98.28}  & 70.27 &  \textbf{94.29} & \textbf{97.91} & 71.18  &  94.86 & \textbf{98.10} & 72.73 & 94.17 &  \textbf{97.68} & 72.40 & 94.84 & \textbf{98.11} & 73.07 & \textbf{95.36} & \textbf{98.31} & 69.44  \\
            BTS-DSN~\cite{Guo2019}   & 95.48 &  98.16  &  71.88 &  94.41  &  97.15 &  69.80 & \textbf{95.02}  & 97.84  &  74.46  &  94.11 &  97.10 & 67.26 & 93.77 & 96.99 & 69.60 &  95.01 &  98.08 & 67.99 \\
            \hline \hline
            LeakGAN 8  & \textbf{95.53}  & 93.30  & \textbf{85.69}  & 94.25 &  96.43 & \textbf{72.65} &  94.35  & 83.21  & \textbf{90.25}  & \textbf{94.28} & 88.25 & \textbf{86.23} & \textbf{94.95} & 89.16 & \textbf{86.85} &  94.68  &  80.30 & \textbf{83.23} \\
            \hline  
    \end{tabular}}
    \label{table:cross_performance}
\end{table*}

\begin{figure}[!thbp]
  \centering
    \includegraphics[width=0.35\textwidth]{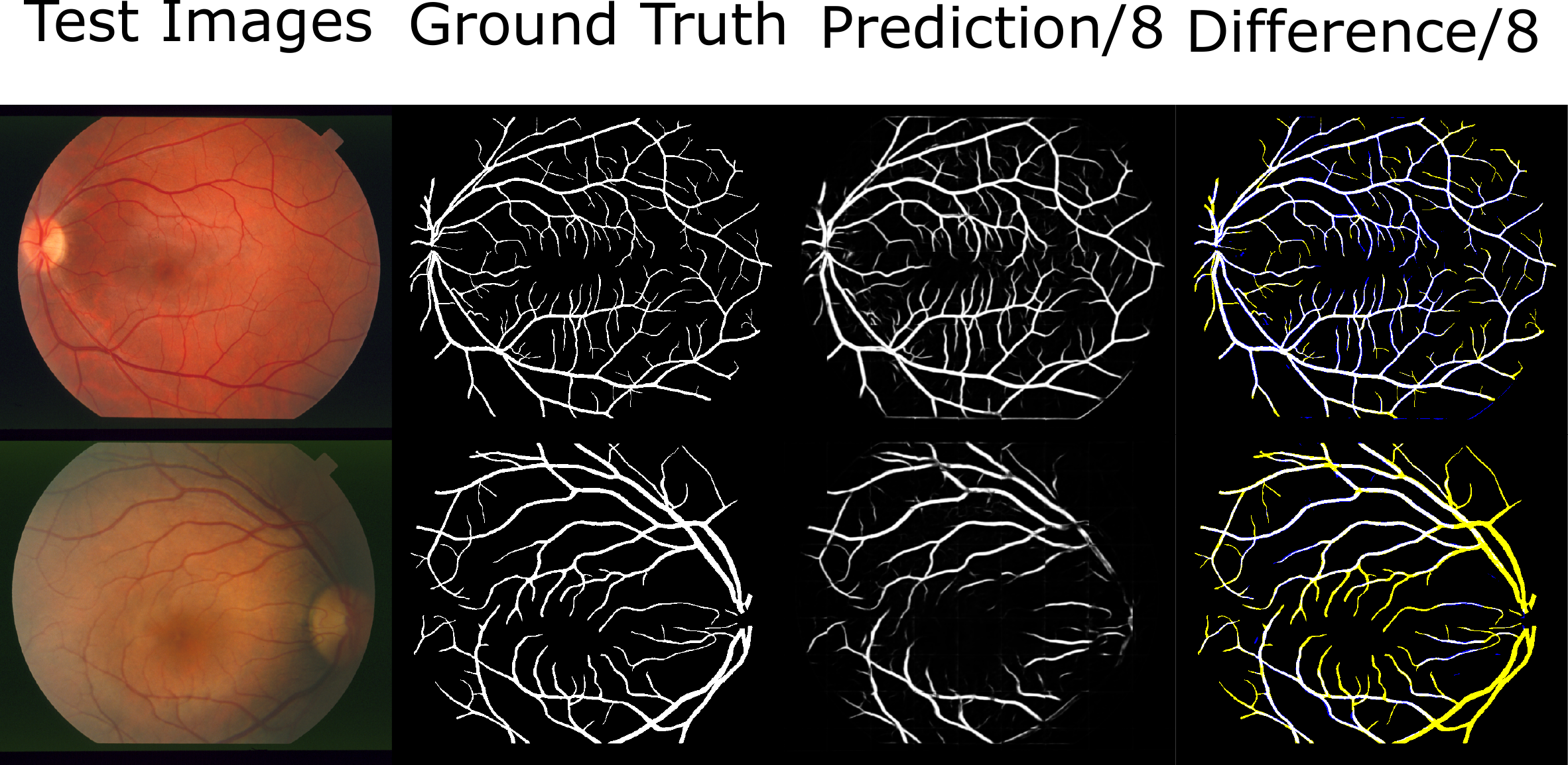}
  \caption{Visualization of the cross-domain prediction results for DRIVE to STRAE. For the difference, the blue pixels indicating false positive, and yellow indicating false negative.}
  \label{fig:vis_cross_domain}
\end{figure}

Fig.~\ref{fig:vis_cross_domain} is the visualization for the evaluations of cross-domain segmentations
from DRIVE to STARE, i.e., DRIVE is the training dataset, and STARE is the test dataset. The STARE vessels' entire structure still keeps clearly though some predictions are weak compared with the ground truth (yellow pixels in the figure). 
We also evaluate our model on the STARE to DRIVE and healthy DRIVE to pathological DRIVE. More detailed results for these different scenarios are shown in Supplementary.

\subsection {Model Analysis}
\label{subsec:ablation}

\begin{table*}[!thbp]
    \centering
    \caption{\small Ablation results under the different labelled and unlabelled settings; L: Labelled, UL: Unlabelled.}
    \resizebox{0.85\textwidth}{!}{
        \begin{tabular}{|l|l|c|c|c|c|c|c||c|c|c|c|c|c|}
            \hline
            \multirow{2}{*}{Methods} & \multirow{2}{*}{Dataset} & \multicolumn{3}{|c|}{(2L, 5UL)} & \multicolumn{3}{|c||}{(5L, 5UL)} & \multicolumn{3}{|c|}{(2L, 6UL)} & \multicolumn{3}{|c|}{(2L, 8UL)}  \\
            \cline{3-14}
            &     & Acc(\%)  & Sp(\%) & Se(\%) & Acc(\%)  & Sp(\%) & Se(\%) &Acc(\%)  & Sp(\%) & Se(\%) &Acc(\%)  & Sp(\%) & Se(\%) \\
            
            \hline \hline
            \multirow{3}{*}{U-Net} & DRIVE & 92.86  & 95.66  & 80.59  &  93.85  & 89.31  &  91.01  & 92.86  & 95.66  & 80.95  & 92.86  & 95.66  & 80.59  \\
            & STARE & 93.16  & 83.67  & 94.60  & 94.87  & 89.74  & 90.03   & 93.16  & 83.67  & 94.60  & 93.16   & 83.67  &  94.60  \\ 
            & CHASE\_DB1 & 94.83  & 95.19  & 83.21  & 95.63  & 92.65  & 88.72  & 94.83  & 95.19  & 83.21  & 94.83  & 95.19  & 83.21     \\ 
                    \hline
            \multirow{3}{*}{U-Net+GAN} & DRIVE &  94.26 & 87.65  & 96.50  & 94.36  & 89.21  & 93.20  &  94.61 & 75.83  & 94.86 & 93.54  & 80.45   &  93.13  \\
            & STARE & 93.98  & 88.51  & 86.17  & 93.86  & 88.23  & 91.58  & 92.32  & 90.97  & 80.93  &  93.71 & 90.06  & 89.63   \\  
            & CHASE\_DB1 & 94.88  & 90.62  & 89.09 & 94.66  & 95.21  & 84.65  & 94.83  & 90.74  & 89.18  & 94.96  & 88.61  &  89.96   \\
                        \hline
            \multirow{3}{*}{U-Net+GAN} & DRIVE &  94.05 & 87.63  & 91.58  & 94.48  & 80.75  & 95.76  &  94.86 & 87.53  & 92.67 & 93.54  & 80.28   &  93.55  \\
            & STARE & 94.08  & 89.21  & 86.06  & 93.78  & 83.94  & 90.74  & 93.57  & 88.64  & 85.85  &  93.86 & 90.04  & 83.08   \\  
            +MT& CHASE\_DB1 & 94.10  & 83.68  & 94.51 & 93.84  & 82.23  & 95.57  &  94.30  & 91.50  & 86.21 & 94.38  & 85.51  &  92.08   \\
                        \hline
            \multirow{3}{*}{U-Net+GAN} & DRIVE & 94.56  & 88.52  & 97.85 & 94.61 & 83.14  & 96.34  & 93.15  & 84.35  & 95.98  & 94.55  & 86.96  & 94.62   \\
            & STARE & 94.32  & 85.17  & 96.31   & 94.89  & 88.21  & 92.32  & 93.73  & 89.74  & 90.23 & 94.05 & 89.31 & 95.12 \\  
            +MT+LM& CHASE\_DB1 & 94.85  &  87.16  & 90.49  & 96.67  & 92.47  & 95.68  & 94.87  & 86.19  & 93.36  & 94.56  & 86.17  &  92.27   \\
            \hline
    \end{tabular}}
    \label{table:lb_unlb}
\end{table*}

\begin{table*}[!thbp]
  \centering 
  \caption{\small Performance of different leaking module setting}
  \resizebox{0.85\textwidth}{!}{
    \begin{tabular}{|c|c|c c c|c|c c c|c|c c c|}
      \hline
      Dataset & Type 1 & Acc(\%) & Sp(\%)  & Se(\%) & Type2  & Acc(\%) & Sp(\%) & Se(\%) &  Type3 & Acc(\%) & Sp(\%)  & Se(\%)  \\
      \hline \hline
      \multirow{3}{*}{DRIVE}  & 1st & 95.72  &  86.72 & 97.50 & 1st+2nd & 94.63  & 80.32 & 97.72  &  \multirow{3}{*}{1st+2nd+3rd} &   &   &  \\ 
      & 2nd    & 93.83 & 83.62 & 96.91  &  1st+3rd & 94.39  & 86.53  & 95.56   &     & 93.58   & 81.40   & 98.18  \\ 
      & 3rd    & 94.30 & 85.48 & 97.12  &  2nd+3rd & 94.69  & 85.79  & 97.61  &   &   &   &    \\ 
      \hline
    \end{tabular}}
  
  \label{table:different_leaking_layers}
\end{table*}

\noindent
\textbf{Ablation studies:} To better understand the contribution made by the modules of the proposed LeakGAN model, we carry out an ablation study for each module. All the experiments are designed under a semi-supervised setting.
\begin{itemize}
    \item U-Net: only labelled images are used for the U-Net training and testing;
    \item U-Net + GAN: GAN-based semi-supervised learning strategy is involved;
    \item U-Net + GAN + MT: Mean-teacher is further added;
    \item U-Net + GAN + MT + LM: Finally, the full model with information leaking involved. 
\end{itemize}

We also experiment with using fewer labelled and unlabelled samples compared with the setting in Table~\ref{table:performance}. From Table~\ref{table:lb_unlb}, we can see that the U-Net can give good performance with few labelled retina images. However, it is prone to  overfit, especially when only two samples are provided. The methods augmented by GAN and further by other modules perform increasingly better. With fixed amount of unlabelled images, more labelled images achieve higher performance. On the other hand, using more unlabelled data also gives higher accuracy for the three datasets. The full framework follows the same trends, but its overall performance across all datasets is the best among all models in comparison.

\noindent
\textbf{Different information leaking settings:} 
According to our model's structure, there are four outputs from the generator that can be concatenated to the upsampling path of discriminator (from high-level features to low-level features) in different ways. The first injection with a shape of $8 \times 8$, then the size is doubled for the following ones up to $64 \times 64$. We evaluate three different leaking outputs (1st, 2nd, and 3rd intermediate outputs of the generator). Table~\ref{table:different_leaking_layers} shows the results with different concatenated style. During the experiments, the $\alpha$ parameter is set to 1. From the table, we can see that the performance for different leaking settings is similar, showing no strong sensitivity to the leaking schemes.


We have also evaluated the effect of using different $\alpha$ values for the leaking module. 
Take the DRIVE experiment as an example as shown in Table~\ref{table:alpha_value}; when $\alpha=1$, the accuracy is the highest. Based on these evaluations suggest that we can choose to leak the first convolutional outputs of the generator and set the $\alpha$ to 1 for the leaking module.  

\begin{table}[H]
    \centering
    \caption{\small Performance of different $\alpha$ values for the leaking module}
    \resizebox{0.38\textwidth}{!}{
        \begin{tabular}{|c|c|c|c|c|c|}
            \hline
            $\alpha$ value & 0.1 & 0.5 & 1.0 & 1.1 & 1.5 \\
            \hline \hline
            Acc(\%)  & 94.13    &  94.20  & 95.72 & 94.61  & 94.67   \\
            Sp(\%)   & 80.54    &  82.48  & 86.72 & 80.13  & 81.47   \\
            Se(\%)   & 98.10    &  98.45  & 97.50 & 97.56  & 98.30   \\
            \hline
    \end{tabular}}
    \label{table:alpha_value}
\end{table}

\noindent
\textbf{Analysis of the focal consistency loss:} To evaluate our focal consistency's performance, we first investigate the different $\rho$ and $\alpha_t$. We also compare it with traditional consistency based on MSE in DRIVE. During the experiments, we follow the protocol in~\cite{Lin2017a}. The results are shown in Table~\ref{table:focal_loss}. We can see that the best performance occurs with $\alpha_t=2.0$ and $\rho=0.25$. 
\begin{table}[H]
  \centering
  \caption{Evaluations of the $\alpha_t$ and $\rho$ of focal loss for DRIVE.}
  \resizebox{0.28\textwidth}{!}{
  \begin{tabular}{|c c|c c c|}
    \hline
    \multicolumn{2}{|c|}{Focal Loss} & \multirow{2}{*}{Acc(\%)}  & \multirow{2}{*}{Sp(\%)}  & \multirow{2}{*}{Se(\%)} \\
    \cline{1-2}
    $\alpha_t$ & $\rho$ &   &   &   \\
    \hline
      1.0 & 0.25   & 92.10 & 83.56 & 93.72  \\
      2.0 & 0.25   & 95.72  & 86.72 & 97.50  \\
      5.0 & 0.25  & 94.69 & 83.84 & 97.53 \\
      \hline
      \multicolumn{2}{|c|}{MSE} & 94.47 & 85.63 & 96.15  \\
      \hline
  \end{tabular}}
  \label{table:focal_loss}
\end{table}

\section{Conclusion} \label{sec:conclusion}

In this paper, we have proposed a novel semi-supervised semantic segmentation method for retina vessel data. To address the issue of lacking labelled data for training a complex semantic segmentation model, a GAN-based framework is employed that integrates information leaking and mean-teacher mechanisms. Information leaking from the generator to the discriminator pollute the generated examples, which helps to improve the generalization performance. The mean-teacher regularization is used to cope with variations in retina images due to changes in imaging conditions. Furthermore, to address the class imbalance problem, a novel focal consistency loss is proposed for the mean-teacher regularization.


The generalization ability of our proposed model has been demonstrated by extensive empirical validations using three widely used retina image datasets. 
We will apply our model to other kinds of medical images in our future work.

{\small
\bibliographystyle{ieee_fullname}
\bibliography{seg}
}

\clearpage

\section{Supplementary of Semi-Supervised Semantic Segmentation of Vessel Images using Leaking Perturbations} \label{sec:Supplementary}


\subsection{Algorithm summary of the proposed model} \label{subsec:algorithm_in_supp}

Our model is summarized in Algo.~\ref{algo:LeakGAN_algorithm}. Furthermore, LeakGAN can be used for the cross-domain scenarios, which the unlabelled target $\mathbf{X}_{\text{ul}}^t$ also is fed into the model, as shown in  Algo.~\ref{algo:LeakGAN_algorithm_cross_domain}.

\begin{algorithm}[!thbp]
  \SetAlgoLined 
  \DontPrintSemicolon
  \KwIn{Source: $\mathbf{X}_l^s, y_l^s, \mathbf{X}_{\text{ul}}^s$}
  \KwResult{Segmentation classifier (discriminator)}
  $\bm \theta_D, \bm \theta_G, \bm \theta_{\text{MT}}$ $\leftarrow$ 
  initialization\;
  \For{iterations of traning}{
    $\mathbf{X}_l^s, y_l^s, \mathbf{X}_{\text{ul}}^s$ $\leftarrow$ sample mini-batch \\
    $\bm z$ $\leftarrow$ sample from $\mathcal{N}(0, I)$ \\
    Generate ${\widetilde{\mathbf{X}}_f}$ by feeding $\bm z$ through generator $G$ \\ 
    $\mathcal{L}_{\text{sup}}, \mathcal{L}_\text{unsup}, \mathcal{L}_{\text{cons-FL}}, \mathcal{L}_\text{adv-}$ $\leftarrow$ calculated by Eq.~(\ref{eq:GAN_based_supervised_objective}), Eq.~(\ref{eq:GAN_based_unsupervised_objective_GAN}), Eq.~(\ref{eq:focal_consistency}) and Eq.~(\ref{eq:LeakGAN_adversarial_objective}) \\
    \For{iterations of inference model updating}{
      $\bm \theta_D \leftarrow$  -- $\Delta_{\bm \theta_D} \mathcal{L}^*$ \\
    }
    \For{iterations of discriminator updating}{
      $\bm \theta_{G}$ $\leftarrow$ --$\Delta_{\theta_{G}} \mathcal{L}_{\text{adv-}}$\\
    }
  }
  \caption{Training of the LeakGAN}
  \label{algo:LeakGAN_algorithm}
\end{algorithm} 

\begin{algorithm}[!thbp]
  \SetAlgoLined 
  \DontPrintSemicolon
  \KwIn{Source: $\mathbf{X}_l^s, y_l^s, \mathbf{X}_{\text{ul}}^s$, Target:$\mathbf{X}_{\text{ul}}^t$}
  \KwResult{Segmentation classifier (discriminator)}
  $\bm \theta_D, \bm \theta_G, \bm \theta_{\text{MT}}$ $\leftarrow$ 
  initialization\;
  \For{iterations of traning}{
    $\mathbf{X}_l^s, y_l^s, \mathbf{X}_{\text{ul}}^s, \mathbf{X}_{\text{ul}}^t$ $\leftarrow$ sample mini-batch \\
    $\bm z$ $\leftarrow$ sample from $\mathcal{N}(0, I)$ \\
    Generate ${\widetilde{\mathbf{X}}_f}$ by feeding $\bm z$ through generator $G$ \\ 
    $\mathcal{L}_{\text{sup}}, \mathcal{L}_\text{unsup}, \mathcal{L}_{\text{cons-FL}}, \mathcal{L}_\text{adv-}$ $\leftarrow$ calculated by Eq.~(\ref{eq:GAN_based_supervised_objective}), Eq.~(\ref{eq:GAN_based_unsupervised_objective_GAN}), Eq.~(\ref{eq:focal_consistency}) and Eq.~(\ref{eq:LeakGAN_adversarial_objective}) \\
    \For{iterations of inference model updating}{
      $\bm \theta_D \leftarrow$  -- $\Delta_{\bm \theta_D} \mathcal{L}^*$ \\
    }
    \For{iterations of discriminator updating}{
      $\bm \theta_{G}$ $\leftarrow$ --$\Delta_{\theta_{G}} \mathcal{L}_{\text{adv-}}$\\
    }
  }
  \caption{Training of the cross-domain LeakGAN}
  \label{algo:LeakGAN_algorithm_cross_domain}
\end{algorithm}

\subsection {Implementation}
\label{subsec:implementation}

Our experiments are based on implementations using TensorFlow~\cite{Abadi2016}. Our model is trained by the patches randomly cropped from the training images. The patch size is $64 \times 64$. The structure of the generator is similar to that of DCGAN~\cite{Radford2016}, which performs well on image generation tasks. A 1-D random normal noise with length 100 is fed to the generator. It passes through a fully connected layer, and the activation is reshaped to $8 \times 8$. There are overall six convolution layers following the fully-connected layer. LeakyReLU is used as the activation function except for the last layer, where tanh() is used instead. Transposed convolution layers with stride 2 are utilized for the upsampling. Batch normalization is added between the convolutional layers. The intermediate outputs of the generator, which have the same frame size as the discriminator's, are used as the leaked information. The discriminator is of U-Net style and has a structure similar to the model proposed in~\cite{Mondal2018}. The kernel of the consecutive convolution layers is 3$\times$3, and max-pooling with step 2 is utilized to downsample the image patches. After four stacked blocks of convolution and downsampling layers, the patches are encoded to the size of 4$\times$4$\times$512. The decoder of the discriminator has a structure symmetric to the encoder. The encoder's output, also the intermediate outputs of the generator, are concatenated to the decoder for the further upsampling. During the experiments, the output of the first convolutional layer of the generator (features with size 8$\times$8) is utilized for the leaking module. It spends around 1.69s, 1.72s, and 4.70s to generate each image prediction in patch-level for DRIVE, STARE, and CHASE\_DB1, respectively, when our model is run on a single NVIDIA GTX1080 GPU.

\subsection{Qualitative results for STARE and CHASE\_DB1} \label{subsec:qualitative_stare_chasedb1}

In this section, more visualization of the experiments on STARE and CHASE\_DB1 are demonstrated.

\begin{figure*}[!thbp]
    \centering
    \begin{subfigure}[t]{\textwidth}
        \centering
        \includegraphics[width=0.75\textwidth]{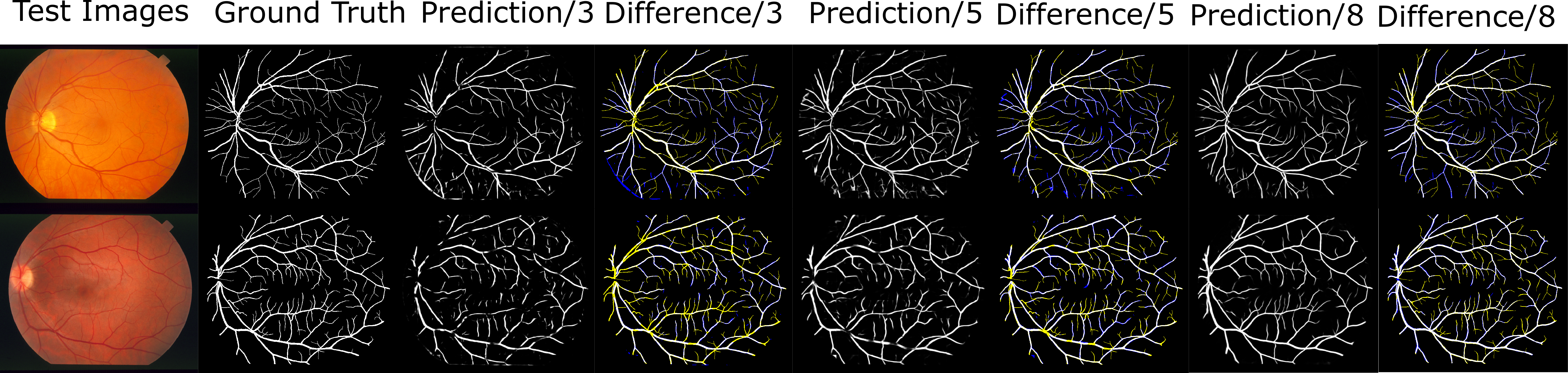}
        \caption{STARE}
        \label{fig:stare_supp}
    \end{subfigure}
    \begin{subfigure}[t]{\textwidth}
        \centering
        \includegraphics[width=0.75\textwidth]{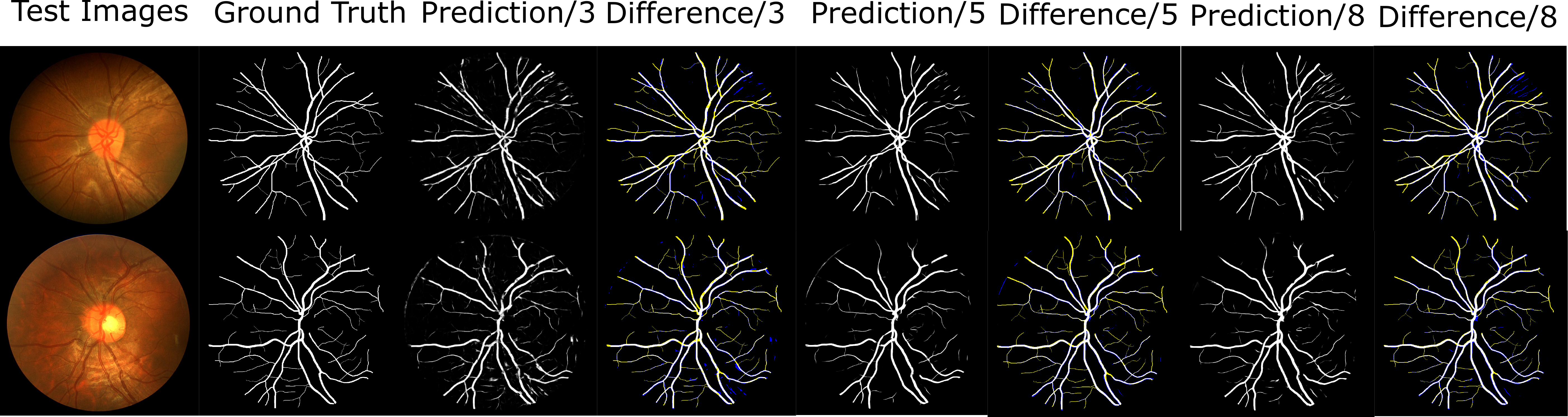}
        \caption{CHASE\_DB1}
        \label{fig:chasedb1_supp}
    \end{subfigure}
    \caption{Visualization of the semantic segmentation on testing images. The first two columns are test images and their ground truth. Then the results are shown when 3, 5 and 8 labelled training images are used. Its differences to ground truth are shown next to the prediction columns, with blue pixels being false positive, and yellow being false negative.}
    \label{fig:vis_entire_supp}
\end{figure*}

The whole-image prediction of the STARE and CHASE\_DB1 are shown in Fig.~\ref{fig:vis_entire_supp}.

\begin{figure*}[!htbp]
    \centering
    \begin{subfigure}[t]{0.5\textwidth}
        \centering
        \includegraphics[width=0.7\textwidth]{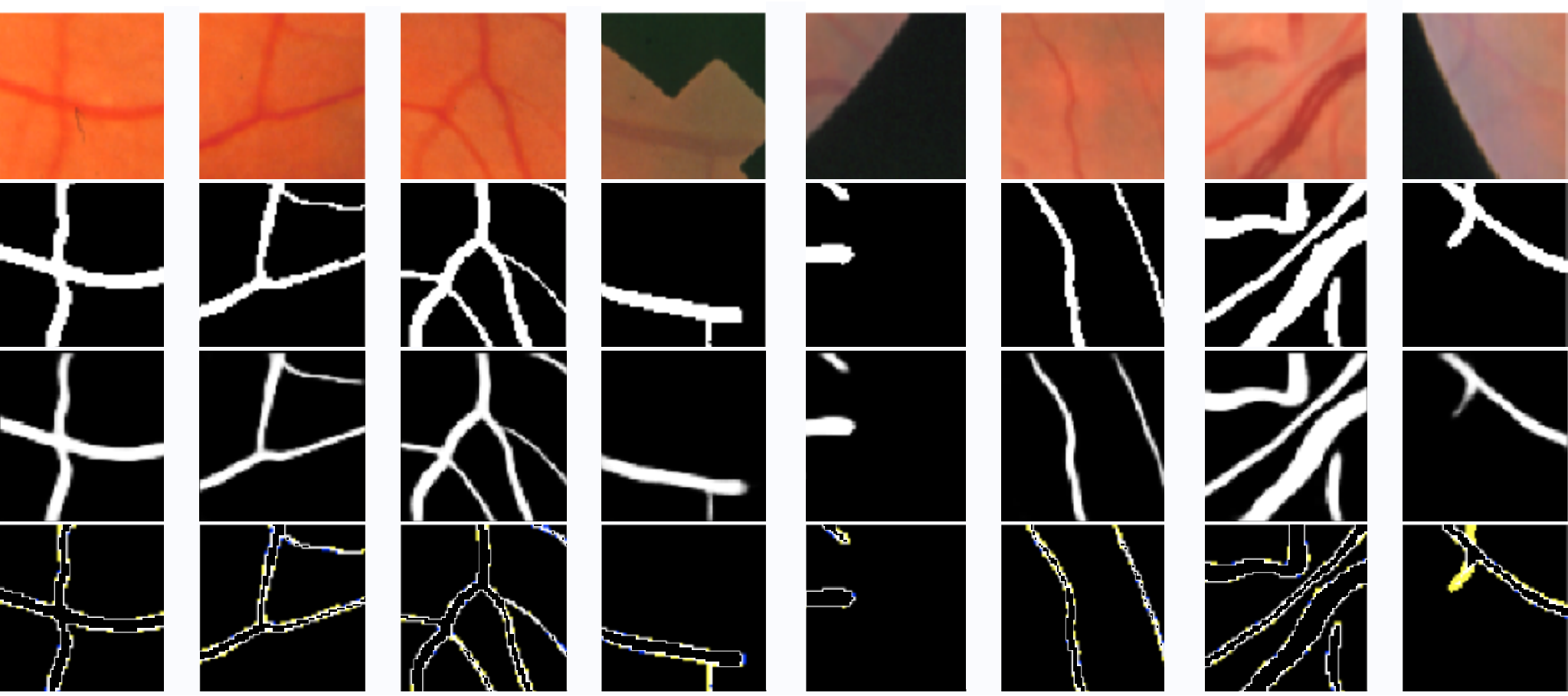}
        \caption{Patch segmentation of STARE}
        \label{fig:stare_patch}
    \end{subfigure}
    \begin{subfigure}[t]{0.5\textwidth}
        \centering
        \includegraphics[width=0.705\textwidth]{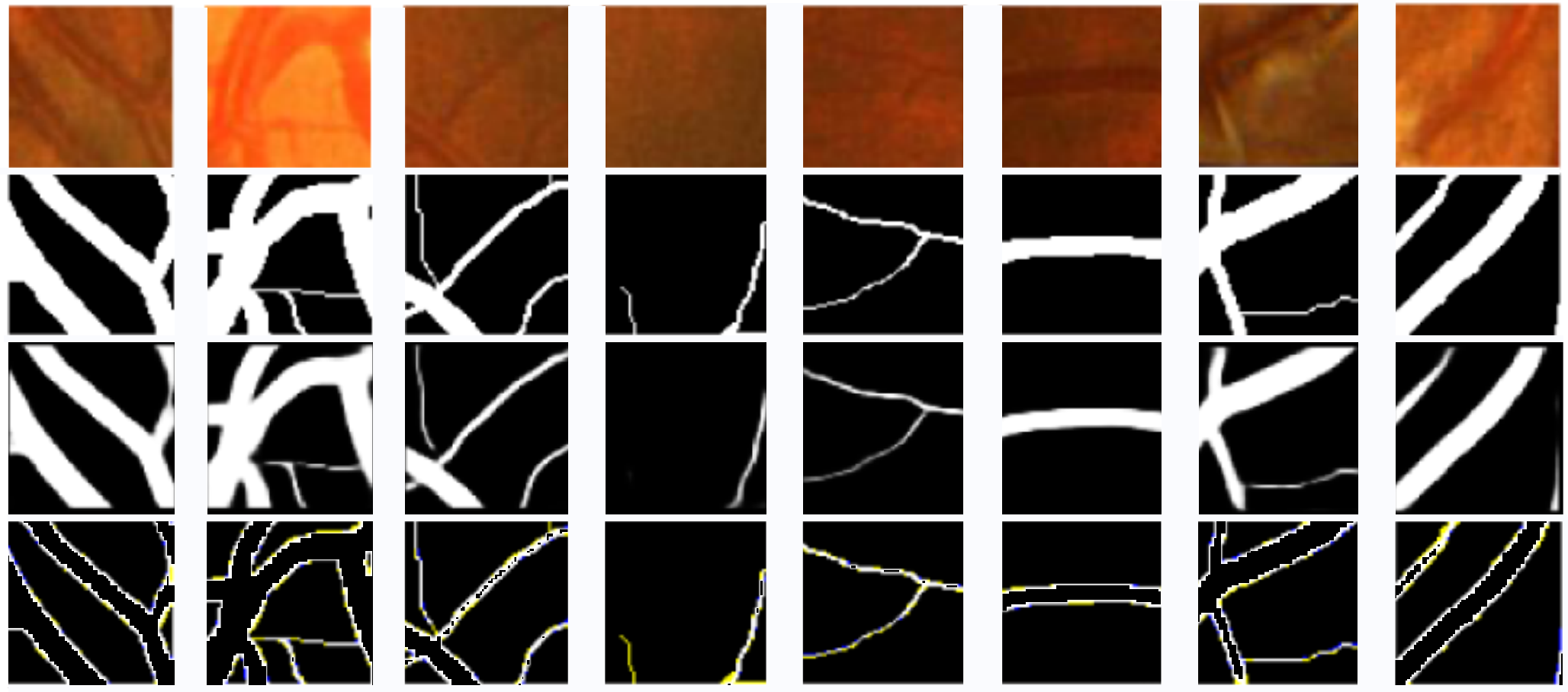}
        \caption{Patch segmentation of CHASE\_DB1}
        \label{fig:chasedb1_patch}
    \end{subfigure}
    \caption{Patched segmentation results for the scenario of 8 labelled training image using different datasets. The testing patches and their ground truth are in first and second row respectively. The third row is for the predictions. The difference between the prediction and ground truth is shown in the last row, blue pixels being false positive, and yellow being false negative.}
    \label{fig:vis_patches_supp}
\end{figure*}

Also, in the sixth column of Fig.~\ref{fig:stare_patch}, the vessels of STARE are detected well though it has some blue pixels around them. However, some vessels' structures disappear (as false negative) in the predictions, e.g. as shown in the fourth column of Fig.~\ref{fig:chasedb1_patch}.

\subsection{More scenarios of cross-domain evaluations} \label{subsec:cross_domain_visualization_supp}

\begin{figure*}[!htbp]
  \centering
  \begin{subfigure}[t]{0.5\textwidth}
    \centering
    \includegraphics[width=0.75\textwidth]{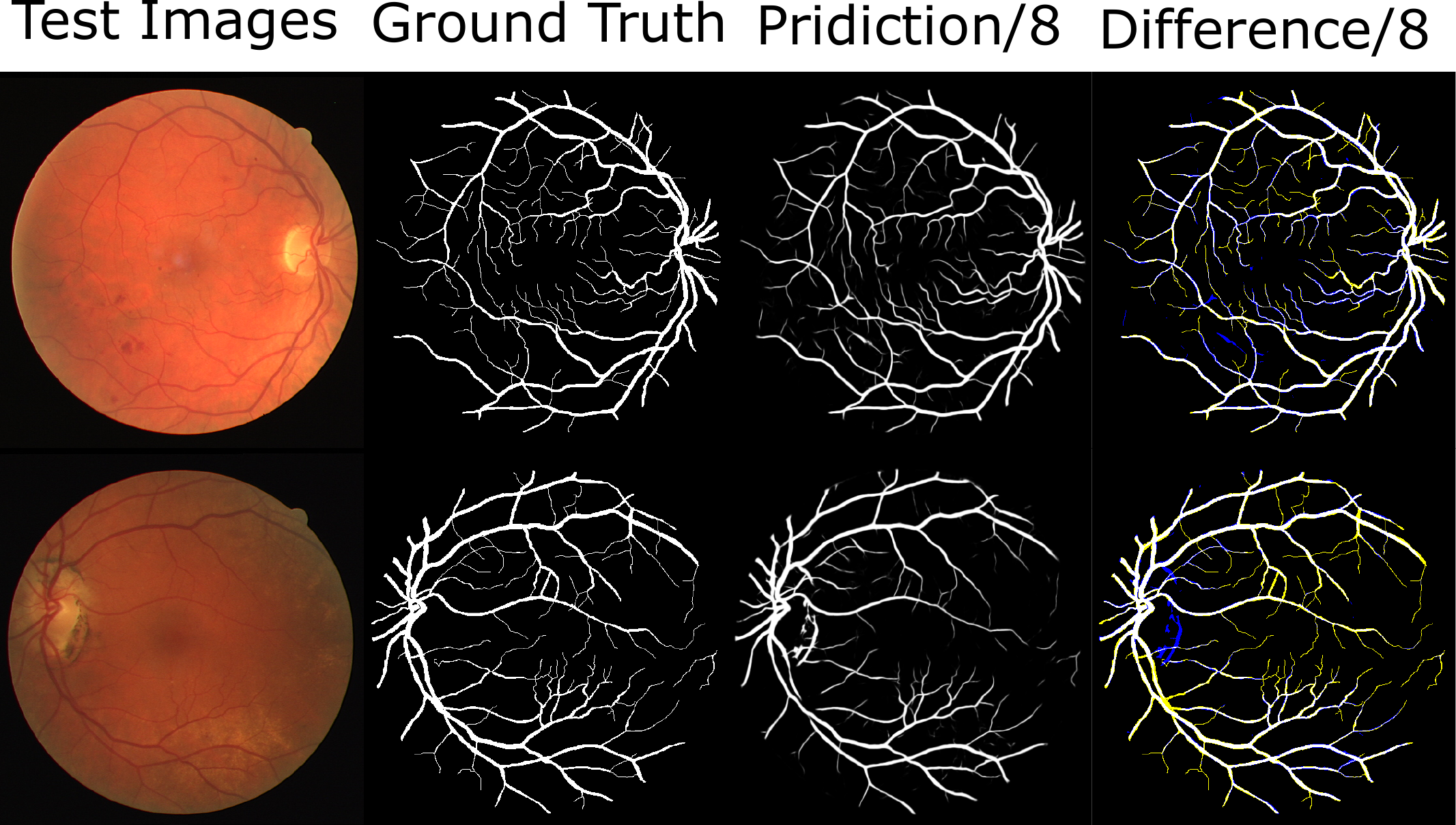}
    \caption{From DRIVE (Healthy) to DRIVE (Diabetic)}
    \label{fig:drive_drive1_cross_supp}
  \end{subfigure}
  \begin{subfigure}[t]{0.5\textwidth}
    \centering
    \includegraphics[width=0.75\textwidth]{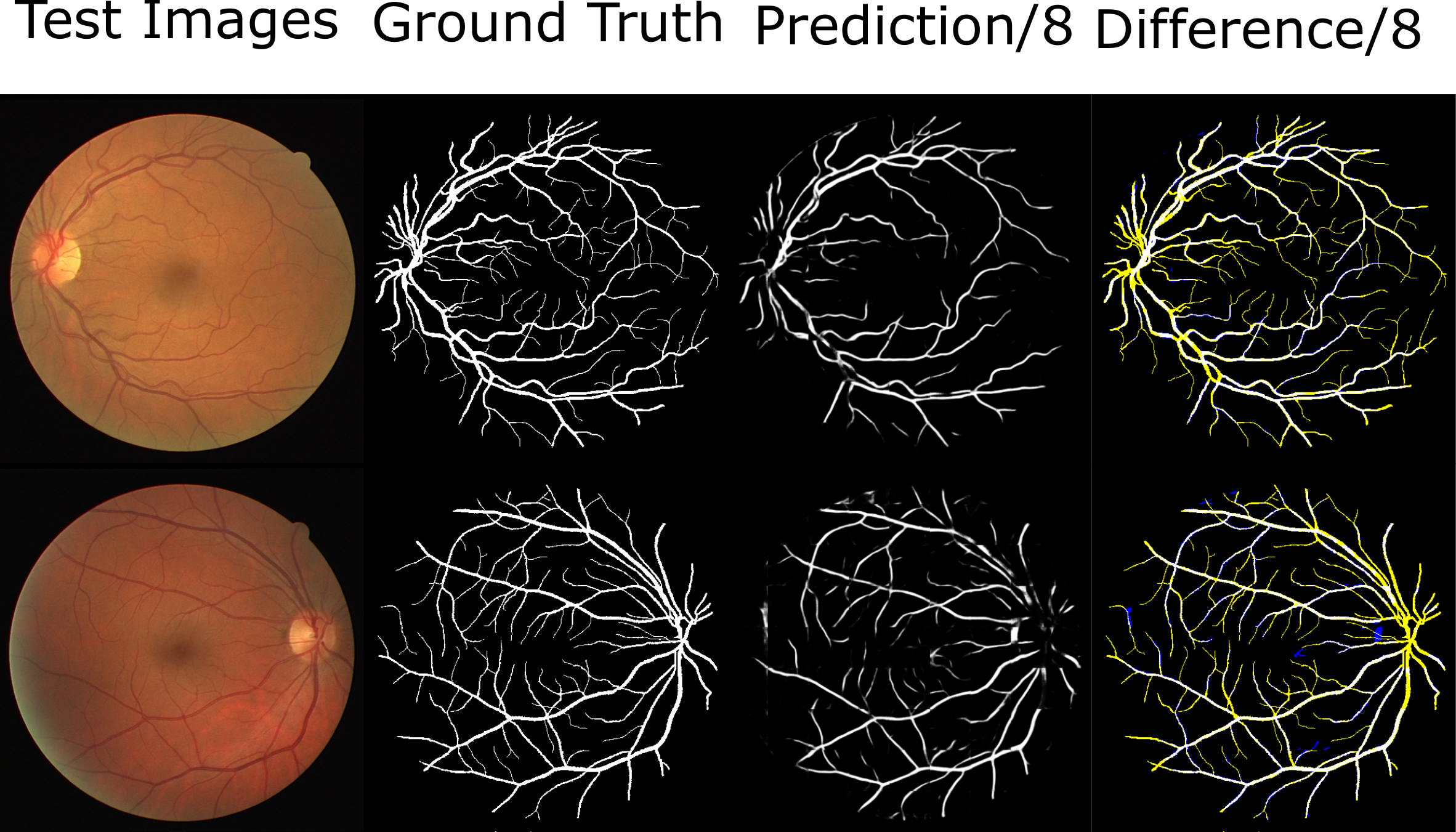}
    \caption{From STARE to DRIVE}
    \label{fig:stare_drive_cross_supp}
  \end{subfigure}

  \caption{Visualization of the cross-domain prediction results. The test images and their ground truth are shown in the first two columns. The predictions are listed on the third. The difference between the prediction and ground-truth is shown in the last column overlaying on ground-truth, with blue pixels indicating false positive, and yellow indicating false negative.}
  \label{fig:vis_cross_domain_supp}
\end{figure*}

We extend the cross-domain segmentation evaluation onto another scenario: training the LeakGAN framework on the healthy images and testing it on the pathological images. 
For the DRIVE dataset, there are seven images from diabetic patients. We also evaluate the performance for the health retina images as training images, the diabetic patient's retina images for the test. 
The results are given in Table~\ref{table:DH_DD}, and segmentation examples are shown in Figure~\ref{fig:drive_drive1_cross_supp}. Also the visualization of the scenario from STARE to DRIVE is shown in Fig.~\ref{fig:stare_drive_cross_supp}. From these results, we can see that the model performs reasonably in this scenario.

Furthermore, we evaluate our model's generalization when it is applied to unseen datasets. During the experiments, the model is trained following  Algorithm~\ref{algo:LeakGAN_algorithm} on one dataset, while the testing is done on retina images of another dataset. For example, these results are obtained: Acc(\%)=95.06, Sp(\%)=92.15, and Se(\%)=83.37, when the transfer from DRIVE (training) to STARE (testing) is considered. This shows that our model also generalizes well to unseen data resources.

\begin{table*}[!thbp]
    \centering
    \caption{\small Performance for training on DRIVE(H), testing on DRIVE(D). Numbers for the training setup are for labelled and unlabelled images respectively.}
    \resizebox{0.6\textwidth}{!}{
        \begin{tabular}{|c|c|c|c|c|}
            \hline
            Train Dataset & DRIVE(H)(5+5) & DRIVE(H)(10+10)  &  DRIVE(H)(10+23)  \\
            Test Dataset & DRIVE(D)  &   DRIVE(D)   &  DRIVE(D) \\
            \hline \hline
            Acc(\%)  & 92.15    &  92.23  & 93.61  \\
            Sp(\%)   & 85.39    &  85.32  & 80.52  \\
            Se(\%)   & 97.53    &  95.16  & 98.30  \\
            \hline
    \end{tabular}}
    \label{table:DH_DD}
\end{table*}

%
%

\end{document}